\documentclass[12pt]{article}

\usepackage{authblk}

\usepackage{hhline}

\usepackage{amsmath}
\usepackage{amssymb}
\usepackage{graphicx}
\usepackage{epsfig}

\newcommand{\ui}{{\mathrm{i}}}

\begin{document}

\title{An improved discretization of Schr\"odinger-like radial equations}

\author{Victor Laliena and Javier Campo}
\affil{Instituto de Ciencia de Materiales de Arag\'on \\
(CSIC -- Universidad de Zaragoza) \\ and \\
Departamento de F\'{\i}sica de la Materia Condensada \\
Universidad de Zaragoza \\
C/Pedro Cerbuna 12, 50009 Zaragoza, Spain}

\date{July 3, 2018}

\maketitle

\begin{abstract}
A new discretization of the radial equations that appear in the solution
of separable second order partial differential equations with some rotational
symmetry (as the Schr\"odinger equation in a central potential) is presented.
It cures a pathology, related to the singular behavior of the radial function 
at the origin, that suffers in some cases the discretization of the
second derivative with respect to the radial coordinate.
This pathology causes an enormous slowing down of the convergence to the 
continuum limit when the two point boundary value problem posed by the radial equation 
is solved as a discrete matrix eigenvalue problem. The proposed discretization is a 
simple solution to that problem. Some illustrative examples are discussed.
\end{abstract}


\vfill\eject

\section{Introduction}
\label{sec:intro}

Many theoretical problems in physics have a rotational symmetry that leads to the 
solution of a radial equation.
Examples include the Schr\"odinger equation in a central potential \cite{LandauQM}, 
heat conduction in a cylinder, potential theory, electromagnetic radiation, wave guides,
acoustics, etc. \cite{SommerfeldPDE}.

Several methods have been developed over the years to solve radial equations. Analytical tools
include the WKB and Born approximations, and series expansions. 
There are also a variety of numerical methods, some based on spectral expansions \cite{Rawitscher05}, 
but the most widely used employ some version of the shooting method, which solves the two point boundary 
value problem posed by the radial equation as a initial value problem in which the derivative at 
the origin is tuned until the appropriate asymptotic behavior at large distances is 
obtained \cite{Morrison62,Killingbeck87}.
The initial boundary value problem is solved in a finite difference scheme with some sophisticated
Runge-Kutta or Numerov algorithm \cite{Sakas05,Fang12,Fang13}.

Methods that solve the boundary value problem as a numerical algebra eigenvalue
problem are also used, since they may have some advantages: 
orthogonality (linear independence)
of the solutions is automatically guaranteed, degeneracy in the case of systems of radial equations
poses no big problem, many eigenvalues and eigenvectors can be obtained at once, etc \cite{Fack86}.
Furthermore, the direct solution of the physical boundary value problem is usually more stable than the
artificial initial value problem. See \cite{Simos99,Vigo05} for reviews of finite difference methods
for radial Schr\"odinger equations.

It is well known that discretized spectral problems in polar coordinates present specific difficulties 
related to the singularity of the equation at the origin and to the definition of an appropriate grid.
They have to be treated carefully in any numerical scheme (see for instance reference \cite{Trefethen00}). 

In this paper we identify a singular behavior of the simplest discretization of the second derivative
in the neighborhood of the origin, which appears in some instances of radial equations, causing an 
enormous slowing down of the convergence to the continuum limit. 
This pathological behavior occurs only in specific cases and has not been discussed previously in the
literature. The authors face this problem when computing the fluctuations around a magnetic skyrmion
\cite{Laliena17} and found a simple and interesting solution: modify the discretization of the
centrifugal potential in a way that exactly compensates the pathological behavior of the second derivative.
This improved discretization ensures the proper behavior of the discretized solution in the neighborhood 
of the origin and accelerates enormously the convergence towards the continuum limit.
No detail about the solution of the radial equation was given in Ref.~\cite{Laliena17}.
The new discretization, which we call the \textit{centrifugal improved discretization}, is interesting
for a broader audience and deserves a separate general treatment.

The outline of the paper is as follows.
In Sec.~\ref{sec:rad} some generalities on the radial equation are reviewed.
The centrifugal improved discretization is introduced in Sec.~\ref{sec:disc} and it is illustrated in two
examples, the two dimensional hydrogen atom and the fluctuations around a magnetic skyrmion, 
in sections~\ref{sec:coul} and~\ref{sec:sk}, respectively. Some final comments are given in the conclusions,
Sec.~\ref{sec:conc}.

\section{Radial equations \label{sec:rad}}

Consider the following system of differential equations:
\begin{equation}
\nabla^2 \psi_i - \sum_jV_{ij} \psi_j + \lambda \psi_i = 0,
\end{equation}
where $V_{ij}$ is a function of the coordinates and 
the indices $i$ and $j$ run from $1$ to $N_\mathrm{e}$.
The above equation can be considered as an eigenvalue equation for the operator
$\nabla^2\delta_{ij}-V_{ij}$. If $V_{ij}$ has a rotational symmetry, either about a central point 
or about an axis, the solutions can be separated into a product of functions that depend on a single 
coordinate. In the case of polar spherical coordinates we have
\begin{equation}
\psi_i(r,\theta,\varphi) = \frac{u_{il}(r)}{r}Y_{lm}(\theta,\varphi), \label{eq:polar}
\end{equation}
where $m$ is an integer, $l$ is a non negative integer, and $Y_{lm}$ is the corresponding
spherical harmonic. For cylindric coordinates the wave functions read
\begin{equation}
\psi_i(r,\varphi,z) = \frac{u_{im}(r,k_z)}{\sqrt{r}}\mathrm{e}^{\ui m\varphi} 
\mathrm{e}^{\ui k_zz}, \label{eq:cyl}
\end{equation}
where $m$ is an integer and the values that the real number $k_z$ can take depend on the boundary 
conditions along the symmetry axis, $\hat{z}$. In any case, the radial equation generically reads
\begin{equation}
u_i^{\prime\prime} - \sum_j \frac{A_{ij}}{r^2} u_j - \sum_j U_{ij} u_j + \lambda u_i = 0,
\label{eq:rad}
\end{equation}
where $\lim_{r\rightarrow 0}r^2U_{ij}(r)=0$, so that $U_{ij}$
is the part of the potential $V_{ij}$ that at $r=0$ is less singular than $1/r^2$, and
$A_{ij}/r^2$ represents the contribution of the centrifugal potential and the $1/r^2$
singularities of $V_{ij}$ (for the case of Schr\"odinger equations we assume that the potentials are 
either regular or transition \cite{Frank71}). 

For the simplest Schr\"odinger equation in a regular central potential, $A_{ij}$ is a $1\times 1$ 
matrix, with $A_{11}=l(l+1)$, where $l$ is a non negative integer, the orbital angular momentum. 
If the potential is axisymmetric
and translationally invariant along the symmetry axis (that is, independent of $z$), or
if the system is confined to move in two dimensions, then $A_{11}=m^2-1/4$, where $m$ is
an integer, the component of the angular momentum along the $\hat{z}$ axis.

The matrix $A_{ij}$ is symmetric and can be diagonalized by an orthogonal transformation.
Let its eigenvalues be denoted by $a_i$, and let $S_{ij}$ be an orthogonal transformation that
diagonalizes $A_{ij}$. By making the change of variables $\tilde{u}_i=\sum_j S_{ij}u_j$,
Eq.~(\ref{eq:rad}) reads
\begin{equation}
\tilde{u}_i^{\prime\prime} - \frac{a_i}{r^2} \tilde{u}_i - \sum_j \tilde{U}_{ij} \tilde{u}_j + 
\lambda \tilde{u}_i = 0,
\label{eq:radDiag}
\end{equation}
where $\tilde{U}=SUS^{-1}$. The definition of $u_i$ in Eqs.~(\ref{eq:polar}) or~(\ref{eq:cyl})
imply that it vanishes as $r\rightarrow 0$, since the wave function has to be finite\footnote{The
finiteness of the wave function at the origin can be relaxed in the case of the Schr{\"o}dinger
equation, requiring only square integrability \protect\cite{Frank71}. Also, in the context of
quantized vortices in superfluids, exact solutions of a nonlinear radial Schr\"odinger equation 
that diverges at the origin have been obtained analytically \protect\cite{Toikka12}.}. 
Obviously, the same is true for $\tilde{u}_i$. Then, for $r\rightarrow 0$ Eq.~(\ref{eq:radDiag})
is dominated by its two first terms and becomes
\begin{equation}
\tilde{u}_i^{\prime\prime} - \frac{a_i}{r^2} \tilde{u}_i = 0.
\label{eq:origin}
\end{equation}
The two independent solutions of the above equation are $\tilde{u}_i=r^{\nu_i}$, 
where $\nu_i$ takes one of the following two values
\begin{equation}
\nu_i = \frac{1}{2}\left(1\pm\sqrt{1+4a_i}\right).
\end{equation}
Only solutions with $\nu_i\geq 1$, for the case of spherical symmetry, Eq.~(\ref{eq:polar})
or with $\nu_i\geq 1/2$, for the case of cylindrical symmetry, Eq.~(\ref{eq:cyl}), are physically
realizable.
For the special case $a_i=-1/4$ the physical solution is $\tilde{u}_i=\sqrt{r}$; 
the second solution, $\tilde{u}_i=\sqrt{r}\ln{r}$, is non physical.
 
\section{The centrifugal improved discretization of the radial equation \label{sec:disc}}

To discretize the radial equation, let us consider a regular mesh with 
points $r_n=n\Delta$, $n=1,2,\ldots N_\mathrm{p}$, where $\Delta$ is the discretization parameter,
and the simplest central difference approximation for the second derivative
\begin{equation}
D^2 u_i(r_n) = \frac{1}{\Delta^2} \left[ u_i(r_n+\Delta) + u_i(r_n-\Delta) - 2 u_i(r_n) \right].
\label{eq:D2}
\end{equation}
With the boundary conditions $u(0)=u((N_\mathrm{p}+1)\Delta)=0$, the second difference operator,
$D^2$, is hermitian and the discrete boundary value problem is an eigenvalue problem of linear 
algebra that can be solved numerically with standard numerical algebra methods. 
The continuum and, in many cases, the infinite volume limit have to be approached, so that 
$\Delta\rightarrow 0$ and, eventually, $N_\mathrm{p}\Delta\rightarrow\infty$.

With the above prescription for the discrete second derivative, the discretization error at 
each point $r$ is of order $u_i^{(4)}(r)\Delta^2$, where $u_i^{(k)}(r)$ is the $k$-th derivative of
$u_i(r)$. Far from the origin the fourth derivative of 
$u_i$ is bounded and the rate of convergence to the continuum limit is high. However, as 
$r\rightarrow 0$ the fourth derivative of $u_i$ diverges if $\nu_i<4$ and it is not an integer, 
and thus the convergence to the continuum limit can be very slow. 
For instance, for $u_i(r)=r^{\nu_i}$ we have $u_i^{(4)}(r)\Delta^2 \sim r^{\nu_i-2}(\Delta/r)^2$
and the relative error, $[(D^2u_i(r_n)-u_i^{(2)}(r_n)]/u_i^{(2)}(r_n)$,
 is of order $(\Delta/r)^2$, which can be very large if $r\sim\Delta$.
Notice that this fact is due to the singular behavior of the radial function at the origin and 
it is not cured by using a higher order discretization scheme. For any scheme, the discretization 
error will be of order $u_i^{(k)}(r)\Delta^{k-2} \sim r^{\nu_i-2}(\Delta/r)^2$,
and the relative error of order $(\Delta/r)^2$.

To analyze this problem, which is confined to a small neigbourhood of the origin,
it is enough to consider Eq.~(\ref{eq:origin}). Remembering that $a_i=\nu_i(\nu_i+1)$,
the finite difference version of Eq.~(\ref{eq:origin}) reads
\begin{equation}
D^2 \tilde{u}_i(r_n) = \frac{\nu_i(\nu_i-1)}{r_n^2} \tilde{u}_i(r_n).
\end{equation}
For small enough $\Delta$ the solution has to be close to the continuum solution, and thus
$\tilde{u}_i(r_n) \sim r_n^{\nu_i}$.
Applying $D^2$ to $r_n^{\nu_i}=(n\Delta)^{\nu_i}$ we obtain
\begin{equation}
D^2r_n^{\nu_i} = W_n(\nu_i) r_n^{\nu_i-2},
\end{equation}
where
\begin{equation}
W_n(\nu) = n^2\left[ \left(1+\frac{1}{n}\right)^\nu + \left(1-\frac{1}{n}\right)^\nu - 2 \right]. 
\label{eq:Wn}
\end{equation}

\begin{figure}[t!]
\centering
\includegraphics[width=0.6\linewidth,angle=0]{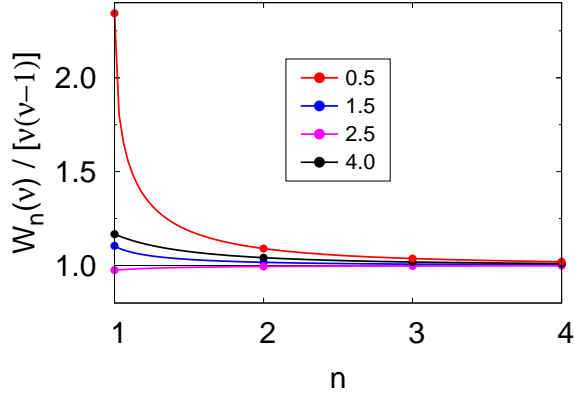}
\caption{The function $W_n(\nu)$, normalized by $\nu(\nu-1)$, as a function of $n$ for the values 
of $\nu$ displayed in the legend. The solid lines represent expression $W_n(\nu)$ given by 
Eq.~(\protect\ref{eq:Wn}) with $n$ extended to the real number set.
\label{fig:Wn}}
\end{figure}

The continuum limit at a given point $r\neq 0$ is obtained as $n\rightarrow\infty$ with 
$\Delta = r/n$. Given that
\begin{equation}
\lim_{n\rightarrow\infty}W_n(\nu) = \nu(\nu-1),
\end{equation}
we see that for any $r>0$ we have $\tilde{u}_i(r_n)\rightarrow r_n^{\nu_i}$ for $\Delta\rightarrow 0$.
However, since in general $W_n(\nu)\neq\nu(\nu+1)$ for small $n$, in the inmediate vicinity of $r=0$
the solution of the discrete problem is very different from $r_n^{\nu_i}$,
no matter how small $\Delta$ is. That is, the asymptotic behaviour of the continuum solution
as $r\rightarrow 0$ is not approximately reproduced in the discrete problem even if $\Delta$ is very 
small.
As we will see in the next sections, this fact slows down dramatically the convergence 
of the spectrum to the continuum limit. 
Fig.~\ref{fig:Wn} displays $W_n(\nu)/[\nu(\nu-1)]$ as a function of $n$ for several values of $\nu$.
Notice that $W_n(\nu)=\nu(\nu-1)$ for $\nu=1$, 2, and 3.

The pathology described above can be easily cured by changing the discretization of the radial 
equation, replacing $a_i$ by $W_n(\nu_i)$ in the term $a_i/r^2$.
This replacement does not affect the continuum limit, and has the virtue of preserving exactly the
continuum asymptotic behavior of $\tilde{u}_i$ as $r\rightarrow 0$ in the discrete version of the
radial equation. That is, the solution of the discretized equation in the neighborhood of $r=0$
exactly reproduces the continuum result, since
\begin{equation}
D^2r_n^{\nu_i} - W_n(\nu_i) r_n^{\nu_i-2} = 0.
\end{equation}

The replacement of the constant coefficient $a_i$ by $W_n(\nu_i)$ in the discrete equation for 
$\tilde{u}_i$ implies that the discrete version of the original radial equation in terms of $u_i$ 
reads
\begin{equation}
D^2 u_i(r_n) - \sum_j \frac{1}{r_n^2}W_n^{ij}u_j(r_n) - \sum_j U_{ij}(r_n) u_j(r_n) + \lambda u_i(r_n) = 0,
\label{eq:imp}
\end{equation}
with
\begin{equation}
W_n^{ij} = \sum_k W_n(\nu_k) v_i^{(k)}v_j^{(k)\,*},
\end{equation}
where the $\vec{v}^{(k)}$, $k=1,\ldots,N_\mathrm{e}$, form a complete orthogonal set of eigenvectors 
of $A_{ij}$. Evidently, the boundary conditions remain unchanged: $u_i(0) = u_i(R) = 0$.

We call the discretization scheme defined by Eq.~(\ref{eq:imp}) the 
\textit{centrifugal improved discretization} (in what follows, improved discretization, to lighten
the writing). The straightforward discretization has the same form, substituting $W_n^{ij}$ by 
$A_{ij}$. In what follows it is called the simple discretization.

\section{The two dimensional hydrogen atom \label{sec:coul}}

As a test we apply the simple and improved discretization prescriptions to a simple exactly 
solvable case, the two dimensional hydrogen atom \cite{Yang91}. It describes the quantum motion 
of an electron constrained to move on a plane under the influence of the Coulomb potential of a 
point nucleus.
Although at first sight it seems to be an artificial model devoid of physical interest, it has been
applied to the study of impurities in very anisotropic solids \cite{Kohn55}. The time independent
Schrodinger equation in cylindric coordinates reads
\begin{equation}
\left[\frac{\partial^2}{\partial r^2}+\frac{1}{r}\frac{\partial}{\partial r}
+\frac{1}{r^2}\frac{\partial^2}{\partial\varphi^2}+\frac{\xi}{r} + k^2\right] \psi = 0
\end{equation}
where $\xi = 2m_\mathrm{e}Ze^2/\hbar^2$ and $k^2=2m_\mathrm{e}E/\hbar^2$, 
with $m_\mathrm{e}$ and $e$ the electron mass and charge, respectively,
$Z$ the nucleus atomic number, and $E$ the energy. Negative and positive $k^2$ correspond to
bound and scattering states, respectively. Notice that the parameter $\xi$ just sets the scale for
spatial variations, and can be eliminated by rescaling the radial coordinate as $\xi r$.
The wave function can be separated into the radial and angular parts as
\begin{equation}
\psi(r,\varphi) = \frac{u(r)}{\sqrt{r}}\mathrm{e}^{\ui m\varphi},
\end{equation}
with $m$ an integer, and the radial equation reads
\begin{equation}
u^{\prime\prime}-\frac{m^2-1/4}{r^2}u+\frac{\xi}{r}u+k^2u = 0. \label{eq:hyd}
\end{equation}
As $r\rightarrow 0$ the physical solution is $u = r^\nu[1 + O(r)]$, with $\nu = 1/2 + |m|$.
The energy of the bound states is given by the formula
\begin{equation}
E_{m,n_\mathrm{r}} = -\frac{E_0}{(|m|+n_\mathrm{r}+1/2)^2}
\end{equation}
where $n_\mathrm{r}$ is a non negative integer, the radial quantum number, and the energy scale is 
set by \mbox{$E_0=Z^2me^4/2\hbar^2$}. Notice that the relation between wave number and energy can 
be cast to the form $E/E_0=4k^2/\xi^2$. Notice also that, as in the three dimensional hydrogen atom, 
the energy levels depend only on the principal quantum number $n=|m|+n_\mathrm{r}+1$, which is a 
positive integer. The radial functions can be expressed in terms of the confluent hypergeometric 
functions \cite{Yang91}.
We are particularly interested in the ground state wave function ($m=0$, $n_\mathrm{r}=0$)
\begin{equation}
u_{0,0}(r) = 2 \xi^{1/2} \sqrt{\xi r}\exp(-\xi r).
\end{equation}

\begin{table}[t]
\centering
\begin{tabular}{cccc}
\hline\hline
\ $m$ \ & \ $n_\mathrm{r}$ \ & \ $\xi R$ \ & \ $\xi\Delta$ \ \\
\hline
0 & 0 & 10 & 0.001 -- 1 \\
0 & 1 & 40 & 0.01 -- 1 \\
1 & 0 & 60 & 0.01 -- 1 \\
\hline 
\end{tabular}
\caption{Parameters used in the computation of several bound states of the two dimensional 
hydrogen atom. \label{tab:sch}}
\end{table}

The radial equations~(\ref{eq:hyd}) have been solved numerically as a numerical linear algebra 
eigenvalue problem using the simple and improved discretizations described in 
Sec.~\ref{sec:disc}. The spectrum was obtained with the help of the ARPACK software package
\cite{ARPACK}. In the remaining of this section the results are briefly discussed.
The parameters used in the computations are gathered in table~\ref{tab:sch}.
As an estimate of convergence rates we fit to the computed energy levels a power law function of 
the grid size $\xi\Delta$, of the form
\begin{equation}
\frac{E^{\mathrm{(num)}}_{\mathrm{m,n_r}}}{E_0} = \frac{E_{\mathrm{m,n_r}}}{E_0} + A (\xi\Delta)^p,
\label{eq:fit1}  
\end{equation}
where $E^{\mathrm{(num)}}_{\mathrm{m,n_r}}$ and $E_{\mathrm{m,n_r}}$ are the numerical and exact energy levels,
respectively, and $A$ and $p$ are the parameters to be fit.
The convergence rate is then quantified by the exponent $p$.

\begin{figure}[t!]
\centering
\includegraphics[width=0.49\linewidth,angle=0]{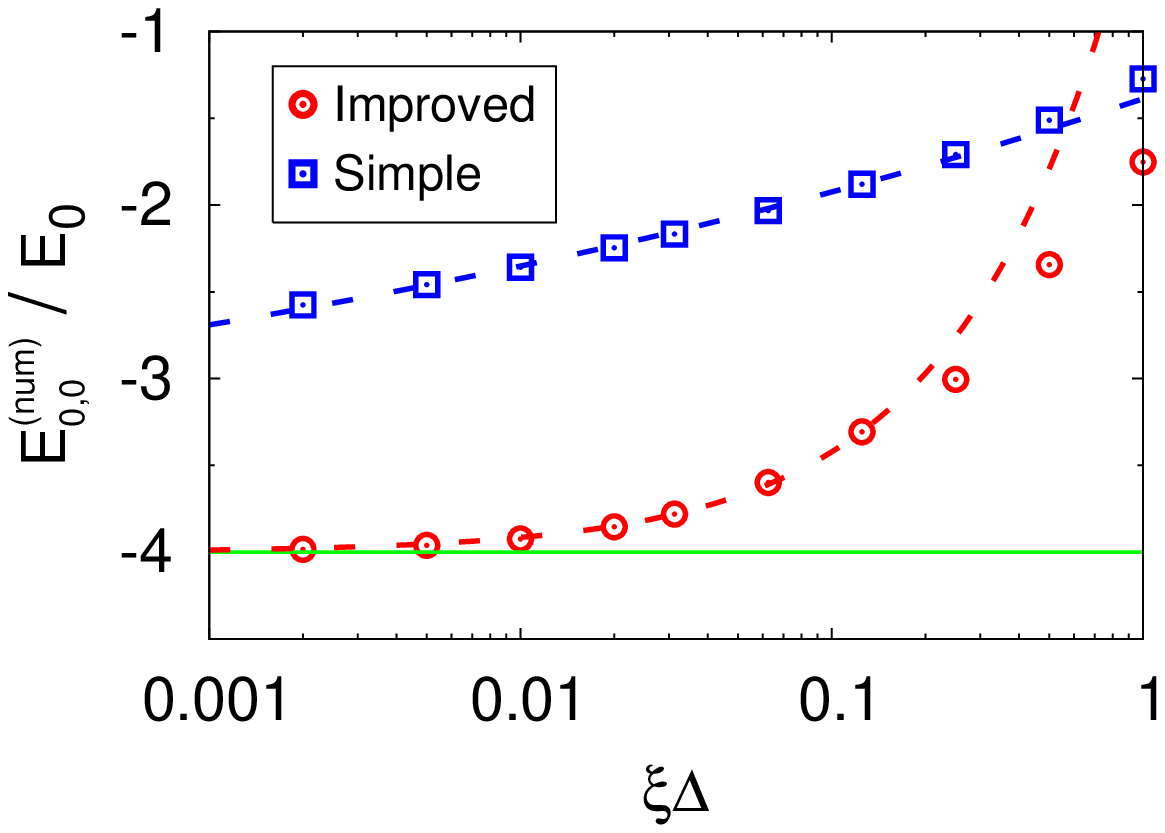}
\includegraphics[width=0.49\linewidth,angle=0]{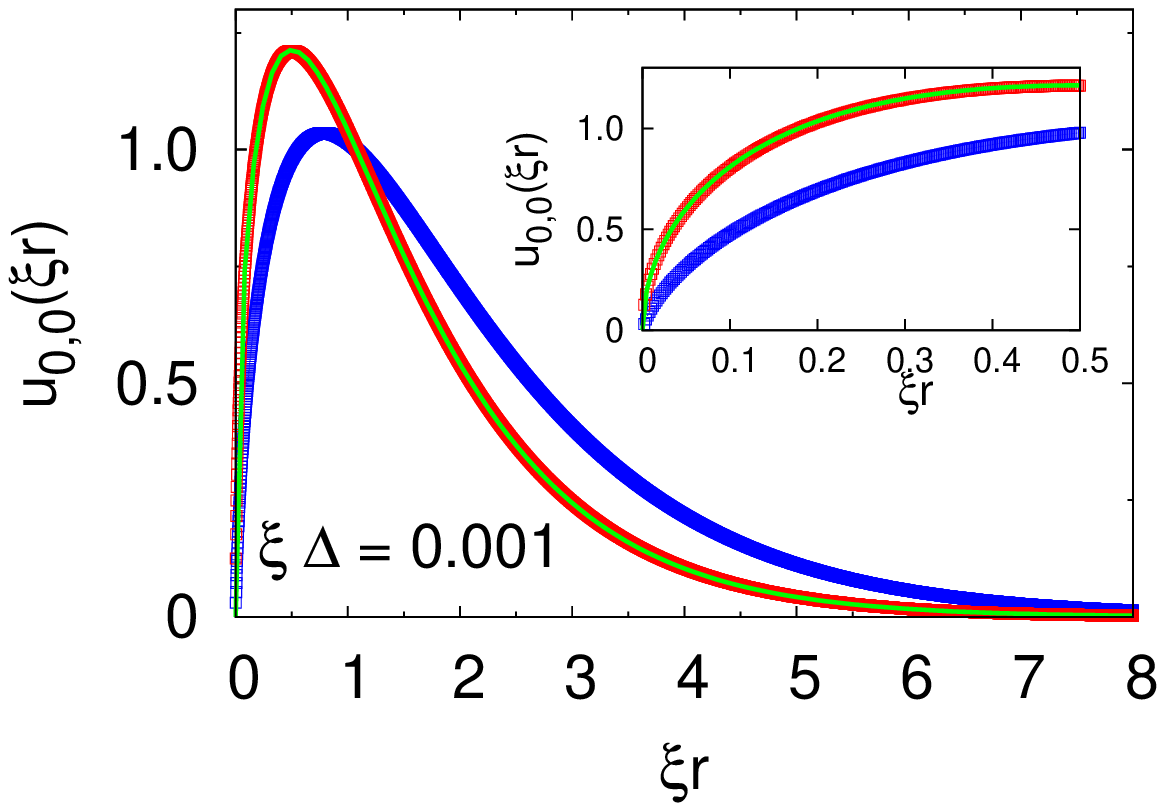}
\caption{Left: energy of the two dimensional hydrogen atom ground state computed numerically
with the simple (blue) and improved (red) discretizations of the radial equation,
as a function of the discretization parameter $\xi\Delta$. The green line is the exact continuum 
result. The blue dashed line represents the function~(\protect\ref{eq:fit1}) with the parameters $A=2.61$ 
and $p=1/10$ obtained via a fit to the data in the range $\xi\Delta\in[0.001,0.02]$;
the red dashed line corresponds to $A=3.94$ and $p=5/6$, obtained via a fit in the 
interval $\xi\Delta\in[0.001,0.2]$.
It is apparent the extremely slow convergence rate of the simple discretization.
Right: ground state radial wave function of the two dimensional hydrogen atom as a function of
$\xi r$, computed numerically with the simple (blue) and improved (red) 
discretizations, with $\xi\Delta =10^{-3}$. The green line is the exact continuum result.
\label{fig:gsHyd}}
\end{figure}

The left panel of Fig.~\ref{fig:gsHyd} displays the lowest lying energy level, 
$E^{\mathrm{(num)}}_{0,0}/E_0$, as a 
function of the discretization parameter $\xi\Delta$. Notice the extremely slow convergence rate
to the continuum limit, with $p=1/10$, as shown by the dashed blue line, which 
corresponds to a fit of Eq.~(\ref{eq:fit1}) to the results in the range 
$\xi\Delta\in[0.001,0.02]$. A fit to the results of the improved discretization, displayed by
the dashed red line in Fig.~\ref{fig:gsHyd} (left), indicates that the convergence rate 
in this case corresponds to $p=5/6$. This is still much slower than the usual second order 
convergence rate, $p=2$. This slowing down can be attributed to the $1/r$ singularity of the 
Coulomb potential.
Nevertheles, we may say that, in comparison with the simple discretization, the convergence rate 
of the improved discretization is extremely fast. 

The right panel of Fig.~\ref{fig:gsHyd} shows the ground state wave function
computed with $\xi\Delta=10^{-3}$ with the simple (blue) and improved (red) discretizations,
and the exact wave function (green line). The inset displays in detail the behavior in the vicinty
of $r=0$. The wave function computed the improved discretization is indistinguisable from the 
exact wave function on the scale of the figure. The simple discretization gives a very inaccurate
ground state wave function, even with this small value of the discretization parameter.
This is due to its behavior as $r^{1/2}$ for $r\rightarrow 0$, which cannot be completely 
reproduced by the simple discretization in the close vicinity of the origin even though 
$\xi\Delta$ is very small.

\begin{figure}[t!]
\centering
\includegraphics[width=0.49\linewidth,angle=0]{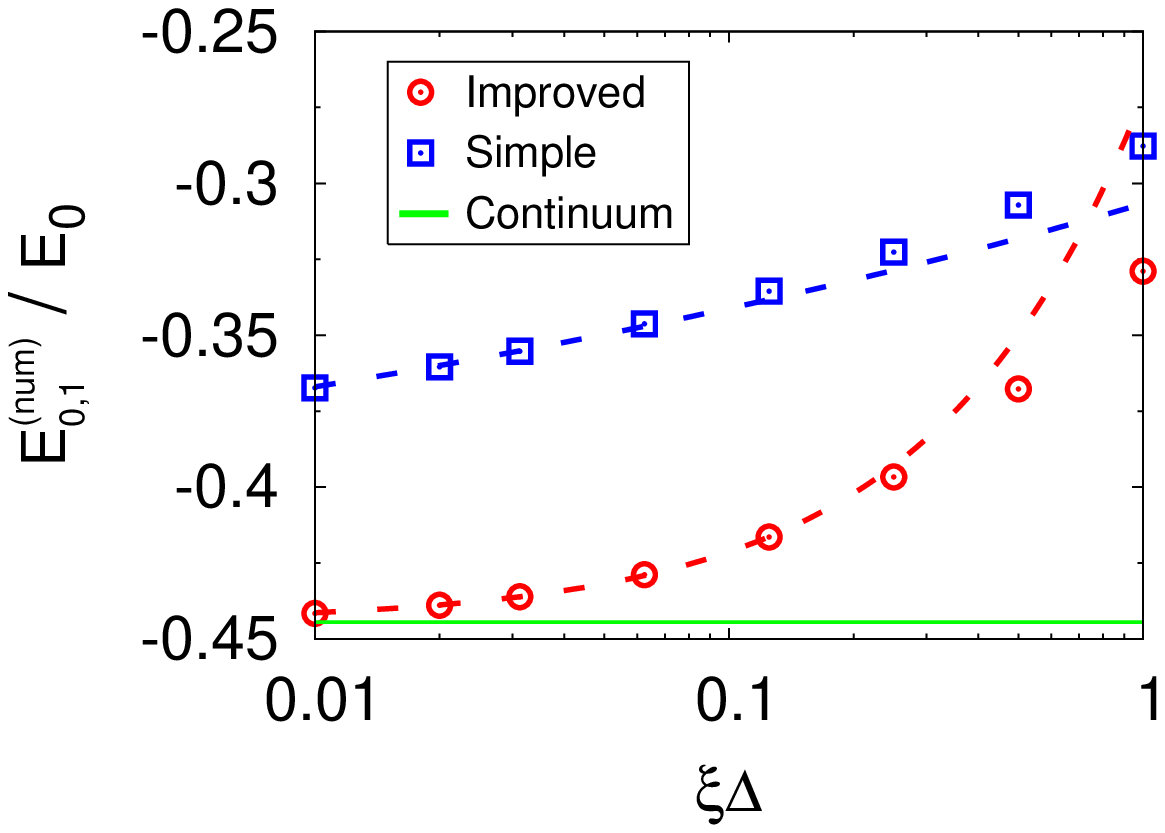}
\includegraphics[width=0.49\linewidth,angle=0]{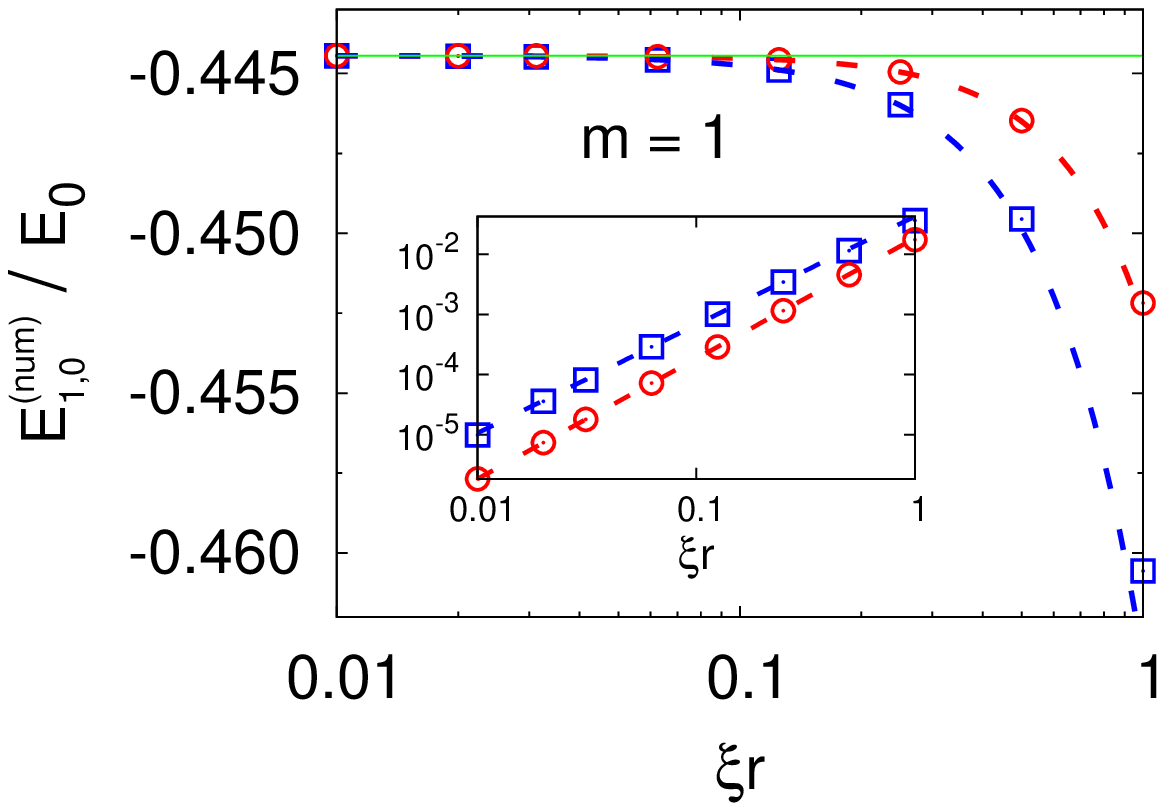}
\caption{Left: first $m=0$ excited state ($n_\mathrm{r}=1$) of the two dimensional hydrogen 
atom computed numerically with the simple (blue) and improved (red) discretizations,
\textit{versus} $\xi\Delta$. The green line is the exact continuum result. 
The blue dashed line corresponds to the fit of~(\protect\ref{eq:fit1}) to the data in the interval 
$\xi\Delta\in[0.01,0.1]$, which gives $p=1/8$;
the red dashed line is the result of a fit to the data in the range $\xi\Delta\in[0.01,0.2]$,
which gives $p=7/8$.
Right: lowest lying energy level with $m=1$ obtained with the simple (blue) and improved (red)
discretizations. 
The dashed lines are fits of the function~(\protect\ref{eq:fit1}) to the data in the range $[0.01,0.1]$, 
which give $p=1.8$ for the simple discretization (blue) and $p=2$ for the improved 
discretization (red).
The inset displays the effect of the discretization, defined as the relative difference between the
result obtained with the corresponding grid and the exact continuum value.
\label{fig:higherHyd}}
\end{figure}

The energy of the first excited state for $m=0$, $E^{\mathrm{(num)}}_{0,1}/E_0$, is displayed as a 
function of $\xi\Delta$ in Fig.~\ref{fig:higherHyd} (left). 
Again the convergence rate is extremely slow in the case of the
simple discretization, since the corresponding wave function behave
as $r^{1/2}$ as $r\rightarrow 0$. Fits of Eq.~(\ref{eq:fit1}) to the results give convergence rates 
with $p=1/8$ and $p=7/8$ for the simple and improved discretizations, respectively. 
Thus, the convergence rate is again much faster with the improved 
discretization than with the simple discretization. The same happens with all states with $m=0$. 

For $|m|>0$ the differences between the improved and simple discretizations are not as dramatic
as in the $m=0$ case. For $|m|=1$ the improved discretization converges substantially faster than 
the simple discretization, as can be seen in Fig.~\ref{fig:higherHyd} (right). 
Fits to the results give convergence rates with $p=1.8$ for the simple discretization and $p=2$ 
for the improved discretization. The inset shows the
relative error introduced by each discretization scheme, $|(E_{1,0}^{\mathrm{(num)}}-E_{1,0})/E_{1,0}|$,
as a function of $\xi\Delta$.

For $|m|>1$ the differences between both discretizations can hardly be noticed.

\section{Fluctuations around a skyrmion magnetic configuration \label{sec:sk}}

Radial equations appear also in the problem of determining the spectrum of fluctuations around 
solitonic field configurations that preserve some rotational symmetry. Examples are the
t'Hooft-Polyakov monopole \cite{tHooft74,Polyakov74}, the Belavin-Polyakov instanton 
\cite{Belavin75,Ivanov07} and the skyrmions \cite{Skyrme61}. Here we consider the solitonic
configurations that appear in two dimensional ferromagnets with isotropic Dzyaloshinskii-Moriya 
interaction, which are generically called skyrmions \cite{Bogdanov94}.

\begin{figure}[t!]
\centering
\includegraphics[width=0.49\linewidth,angle=0]{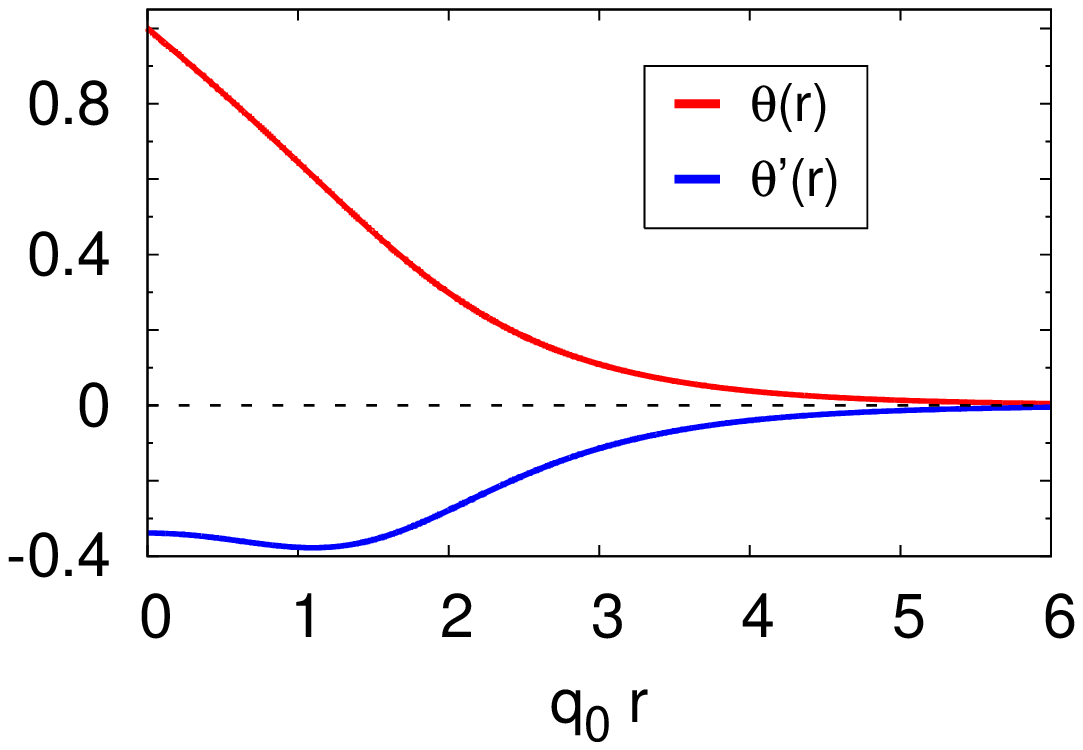}
\includegraphics[width=0.49\linewidth,angle=0]{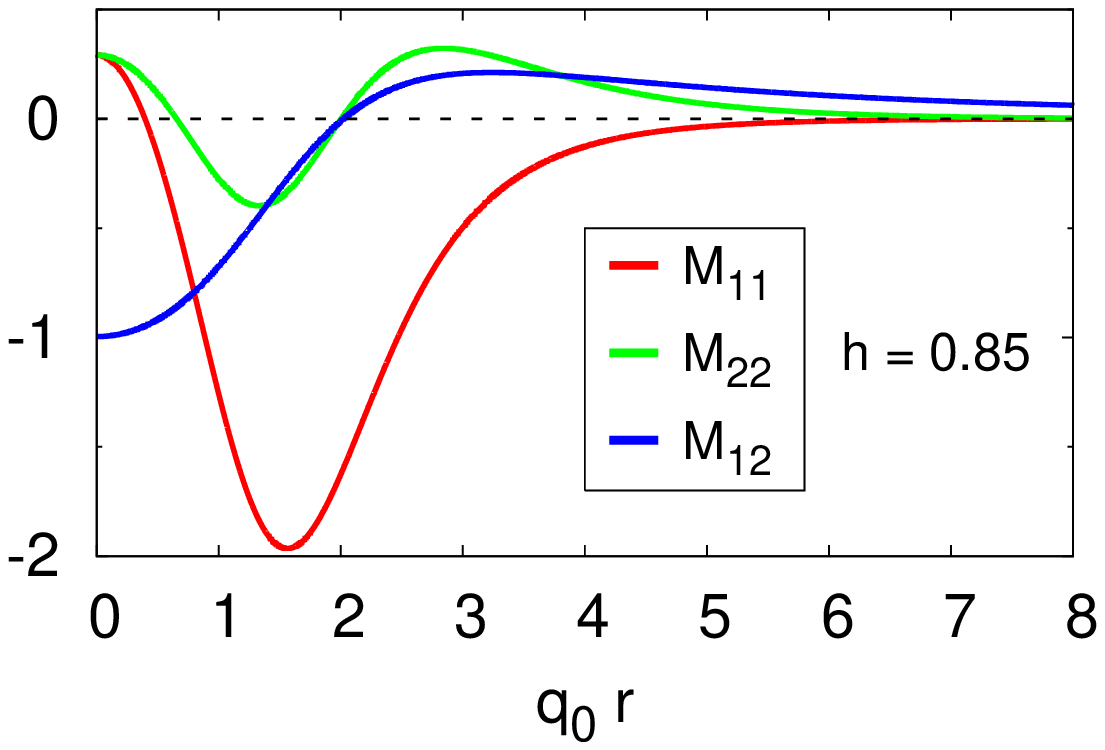}
\caption{Left: the solitonic profile $\theta(r)$ of the magnetic skyrmion and its derivative,
$\theta^\prime(r)$, for $h=0.85$. Right: the matrix elements $M_{\alpha\beta}(r)$ entering the skyrmion
fluctuation operator for $h=0.85$.
\label{fig:theta}}
\end{figure}

Those magnetic skyrmions are two dimensional magnetic configurations represented by a unit vector 
field $\hat{n}(\vec{r})$ that point in the direction of the local magnetic moment, with
unit topological charge and rotational symmetry around the magnetic field.
They are stationary points of the following dimensionless energy functional
\begin{equation}
\mathcal{W} = \int d^2r \left(\frac{1}{2}\sum_i\partial_i\hat{n}\cdot\partial_i\hat{n}
+q_0\hat{n}\cdot\nabla\times\hat{n}-q_0^2\vec{h}\cdot\hat{n}\right). \qquad \label{eq:W}
\end{equation}
where $\partial_i=\partial/\partial x_i$, with $x_i\in\{x,y,z\}$ being the cartesian
coordinates,
$q_0$ is a constant with the dimensions of inverse length,
\begin{equation}
\hat{n}\cdot\nabla\times\hat{n} = 2n_z(\partial_xn_y-\partial_yn_x),
\end{equation}
and $\vec{h}=h\hat{z}$ is the applied magnetic field, perpendicular to the plane on which the 
magnetic system is confined.
The skyrmion is then a solution of the Euler-Lagrange equations corresponding to the
$\mathcal{W}$ functional that, using polar coordinates $(r,\varphi)$ in the plane,
can be parametrized by a single function of the radial coordinate, $\theta(r)$,
as \cite{Bogdanov94,Laliena17}
\begin{equation}
\hat{n}_\mathrm{S}(r,\varphi)=(\sin\theta(r)\cos\varphi,-\sin\theta(r)\sin\varphi,\cos\theta(r))
\end{equation}
with the boundary conditions $\theta(0)=\pi$ and $\lim_{r\rightarrow\infty}\theta(r) = 0$. 
Thus the skyrmion is a solitonic configuration on a ferromagnetic background that has in its center
a magnetic moment opposite to the magnetic field. For $r\rightarrow\infty$, $\theta(r)$ vanishes
asymptotically as $\exp(-\sqrt{h}q_0r)/\sqrt{q_0 r}$.
The function $\theta(r)$ is displayed as a function of $q_0r$ in Fig.~\ref{fig:theta} (left)
for $h=0.85$.

The fluctuations around the skyrmion configuration can be parametrized by two real fields,
$\xi_\alpha$, with $\alpha=1,2$, writing
\begin{equation}
\hat{n} = \left(1-\sum_\alpha\xi_\alpha^2\right)^{1/2}\hat{n}_\mathrm{S} + \sum_\alpha\xi_\alpha\hat{e}_\alpha,
\label{eq:n}
\end{equation}
where $\{\hat{e}_1,\hat{e}_2,\hat{n}_\mathrm{S}\}$ form a right-handed orthonormal triad.
Plugging (\ref{eq:n}) into (\ref{eq:W}) and expanding $\mathcal{W}$ in powers of $\xi_\alpha$,
we get up to quadratic terms (the linear term vanishes on account of the Euler-Lagrange equations)
\begin{equation}
\mathcal{W} = \mathcal{W}(\hat{n}_\mathrm{S}) + \int d^2r \sum_{\alpha\beta}\xi_\alpha K_{\alpha\beta}\xi_\beta
+ O(\xi^3),
\end{equation}
where $K_{\alpha\beta}$ is a differential operator that depends on $\theta(r)$, whose explicit form
is given in \cite{Laliena17}.

\begin{figure}[t!]
\centering
\includegraphics[width=0.65\linewidth,angle=0]{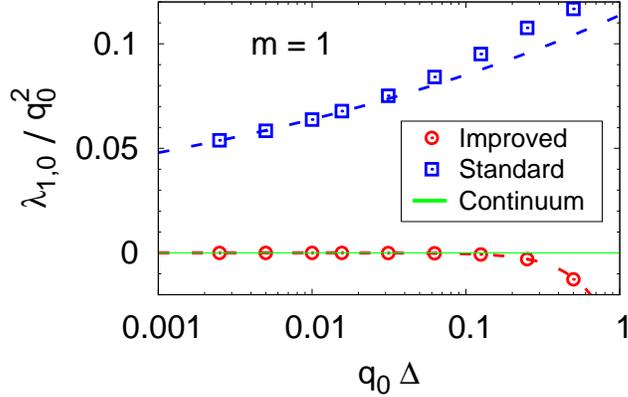}
\caption{
The eigenvalue of the skyrmion fluctuation operator corresponding to the translational zero mode
($m=1$) as a function of $q_0\Delta$, computed with the simple (blue) and improved (red)
discretizations. The convergence rate of the simple discretization is extremely slow: the fit 
of~(\protect\ref{eq:fit2}), with $\lambda^{\mathrm{(c)}}=0$, to the data in the range 
$q_0\Delta\in[0.0025,0.02]$,
represented by the blue dashed line, gives $p=1/8$.   
In contrast, the improved discretization shows a second order convergence rate ($p=2$), which
is the result from the fit to the data (red dashed line).
\label{fig:sk_zeroMode_eval}}
\end{figure}

The spectrum of fluctuations around the magnetic skyrmion is determined by the spectrum of
$K_{\alpha\beta}$. The rotational symmetry around the magnetic field direction implies that the
eigenfunctions of $K_{\alpha\beta}$ can be chosen of the form
\begin{equation}
\xi_\alpha^{(m)}(r,\varphi) = \frac{u_\alpha^{(m)}(r)}{\sqrt{r}} \exp(\ui m \varphi),
\end{equation}
If $\lambda_m$ is the corresponding eigenvalue of $K_{\alpha\beta}$, the radial equations read
\begin{equation}
-u_\alpha^{\prime\prime} + \frac{m^2+3/4}{r^2}u_\alpha
-\mathrm{i}\frac{2m}{r^2} \sum_\beta\epsilon_{\alpha\beta}u_\beta + hq_0^2 u_\alpha
+ \sum_\beta M_{\alpha\beta}u_\beta=\lambda u_\alpha, \label{eq:skrad}
\end{equation}
where $\epsilon_{\alpha\beta}$ is the two dimensional antisymmetric tensor and, to avoid cumbersome 
notation, the index $m$ in $u_\alpha$ and $\lambda$ is not explicitely shown.
The matrix $M_{\alpha\beta}$, which is analytic in $r=0$ and vanishes exponentially as 
$r\rightarrow\infty$, is 
\begin{eqnarray}
M_{11} &=& -2\frac{\sin^2\theta}{r^2}-2q_0\frac{\sin(2\theta)}{r} + hq_0^2 (\cos\theta-1), \\
M_{22} &=& -\frac{\sin^2\theta}{r^2}-q_0\frac{\sin(2\theta)}{r} -\theta^\prime(\theta^\prime+2q_0)
+ hq_0^2 (\cos\theta-1), \\
M_{12} &=& \mathrm{i}2m\left(\frac{1+\cos\theta}{r^2}-q_0\frac{\sin\theta}{r}\right),
\end{eqnarray}
and $M_{21}=-M_{12}$. The boundary conditions are $u_\alpha(0)=0$ and, for bound states, $u_\alpha(R)=0$,
with $R\rightarrow\infty$. For scattering states the boundary condition is related to the oscillatory
behavior of the wave function as $r\rightarrow\infty$, although we may consider the system enclosed 
in a box of radius $R$, so that $u_\alpha(R)=0$, and let $R\rightarrow\infty$.
Notice that $q_0$ merely sets the scale of spatial variations and can be eliminated by a rescaling
of the radial coordinate. Hence, the eigenfunctions depend on $q_0r$ and the eigenvalues are 
proportional to $q_0^2$. The functions $M_{\alpha\beta}$ are displayed in Fig.~\ref{fig:theta} (right),
as a function of $q_0r$ for $h=0.85$.

The asymptotic form of Eq.~(\ref{eq:skrad}) for $r\rightarrow 0$ gives the matrix $A_{ij}$:
\begin{equation}
A = \left(\begin{array}{cc}
m^2 + 3/4 & -\ui 2 m \\
\ui 2 m & m^2 + 3/4 
\end{array}
\right).
\end{equation}
Its eigenvalues are $a_\pm = (m\pm1)^2-1/4$, and therefore the behaviour of the physical
solution $u_\alpha$ (regular at $r=0$) as $r\rightarrow 0$ is given by
\begin{equation}
\nu_\pm = 1/2 + |m\pm 1|.
\end{equation}

Due to the translational invariance of the skyrmion, the operator $K_{\alpha\beta}$ has two independent
normalizable (bound states) zero modes, which have $m=\pm 1$. For the case $m=+1$, its two 
components are
\begin{equation}
\begin{array}{ll}
u_1(r) = -\sqrt{r}\theta^\prime(r), \\
 \\
u_2(r) = -\sin\theta(r)/\sqrt{r}.
\end{array}
\label{eq:zeroMode}
\end{equation}
The discretization breaks the continuous translational symmetry and the discretized operator
does not have exact zero modes: the eigenvalues of the bound states with $m=\pm 1$ tend to zero
in the continuum limit.

\begin{figure}[t!]
\centering
\includegraphics[width=0.49\linewidth,angle=0]{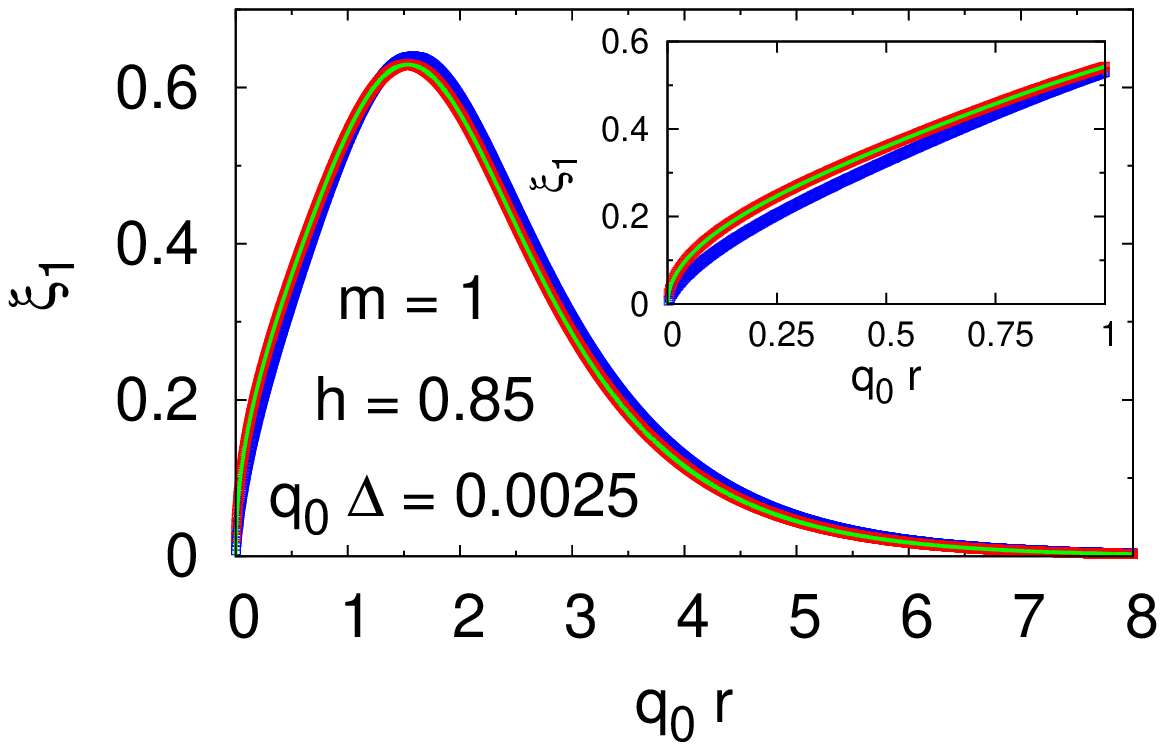}
\includegraphics[width=0.49\linewidth,angle=0]{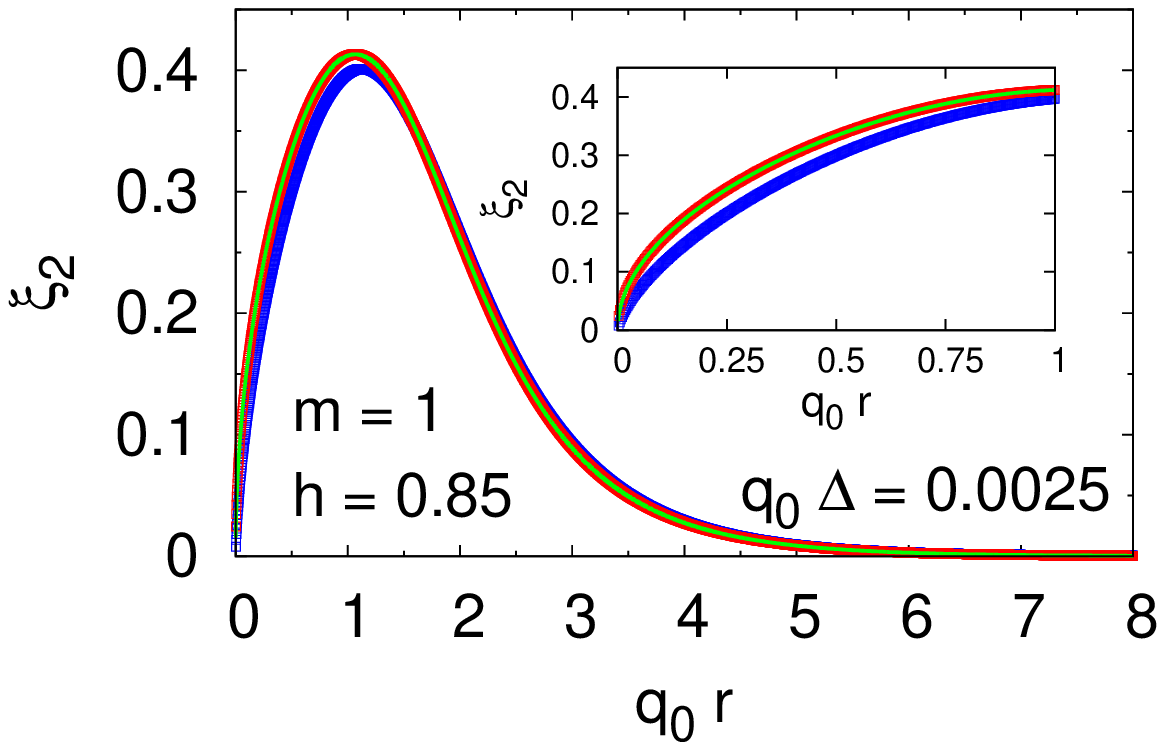}
\caption{Components of skyrmion fluctuation zero mode eigenfunction ($m=1$) 
for $h=0.85$, obtained with the improved (red) and simple (blue) 
discretizations, with $q_0\Delta=0.0025$. The green line is the exact continuum result.
The insets show magnifications of the $r=0$ vicinity.
\label{fig:sk_zeroMode_evec}}
\end{figure}

The continuum spectrum has a gap of magnitude $q_0^2h$, so that it appears for $\lambda\geq q_0^2h$.
The discrete spectrum, which depends qualitatively on $h$, is located below the gap, 
$\lambda<q_0^2h$. 
In addition to the zero mode, there is one bound state with $m=0$ (a breathing mode) for 
any $h$. For $h<0.98$ two (degenerate) bound states with $m=\pm 2$ appear. 
Succesive bound states with correlative higher values of $|m|$ arise by lowering $h$. For
$h<0.57$ the eigenvalue of the $|m|=2$ bound states is negative and therefore the skyrmion
is unstable.

The bound states have been computed by numerical diagonalization, with the ARPACK software 
package \cite{ARPACK}, of the operator~(\ref{eq:skrad}) discretized with the simple and improved 
schemes. To estimate the convergence rates we fit to the computed eigenvalues a power law function of 
the grid size $q_0\Delta$, of the form
\begin{equation}
\lambda/q_0^2 = \lambda^{\mathrm{(c)}} + A (q_0\Delta)^p,
\label{eq:fit2}  
\end{equation}
where $\lambda^{\mathrm{(c)}}$, $A$, and $p$ are the parameters to be fit. 
Evidently, $\lambda^{\mathrm{(c)}}$ is an extrapolation of the results to the continuum limit.
The convergence rate is quantified by the exponent $p$.

Let us discuss the results for the typical case $h=0.85$, for which we set $q_0R=25$.
Fig.~\ref{fig:sk_zeroMode_eval} displays the computed eigenvalue of the $m=1$ bound state, which 
becomes a zero mode in the continuum limit, as a function of the discretization parameter 
$q_0\Delta$. Notice the extremely slow convergence rate in the case of the simple discretization: a 
fit of the function~(\ref{eq:fit2}), with $\lambda^{\mathrm{(c)}}=0$, to the data gives $p=1/8$. 
It is represented by the blue dashed line. 
In comparison, the convergence rate with the improved discretization is extremly fast: the fit
gives $p=2$. The fact that the convergence rate of the improved discretization in 
the skyrmion problem is of the expected second order contrasts with the much slower convergence 
rate in the case of the 2D hydrogen atom. This is likely due to the fact that the potential of the 
skyrmion fluctuations is analytic at $r=0$, while it is coulombian ($1/r$) in the case of the 
hydrogen atom.

\begin{table}[t]
\centering
\begin{tabular}{|c||ccc||ccc|}
\cline{2-7}\cline{2-7}
\multicolumn{1}{c|}{} & \multicolumn{3}{|c||}{Simple} & \multicolumn{3}{c|}{Improved} \\
\hhline{-======}
\ $m$ \ & \ $\lambda^{\mathrm{(c)}}$ \ & \ $A$ \ & \ $p$ \ & \ $\lambda^{\mathrm{(c)}}$ \ & \ $A$ \ & \ $p$ \ \\
\hline
0 & 0.3561 & -0.101 & 2 & 0.3561 & -0.057 & 2 \\
1 &   --   & 0.113  & 1/8 & -- & -0.048 & 2 \\
2 & 0.5987 & -0.232 & 2 & 0.5987 & -0.100 & 2 \\
\hline 
\end{tabular}
\caption{Results of the fits of the function~(\protect\ref{eq:fit2}) to the numerical results for the
bound state spectrum of skyrmion fluctuations with $h=0.85$. 
\label{tab:fit_sk}}
\end{table}

The components of the zero mode eigenfunction with $m=1$ are displayed in 
Fig.~\ref{fig:sk_zeroMode_evec} for $q_0\Delta=0.0025$. 
The insets are a magnification of the vicinity of $r=0$.
The green lines are the exact continuum results given by Eqs.~(\ref{eq:zeroMode}). 
In the scale of the figure they are indistinguisable from the eigenfunction components
computed with the improved discretization; in the case of the simple discretization there are 
noticeably differences.

\begin{figure}[t!]
\centering
\includegraphics[width=0.49\linewidth,angle=0]{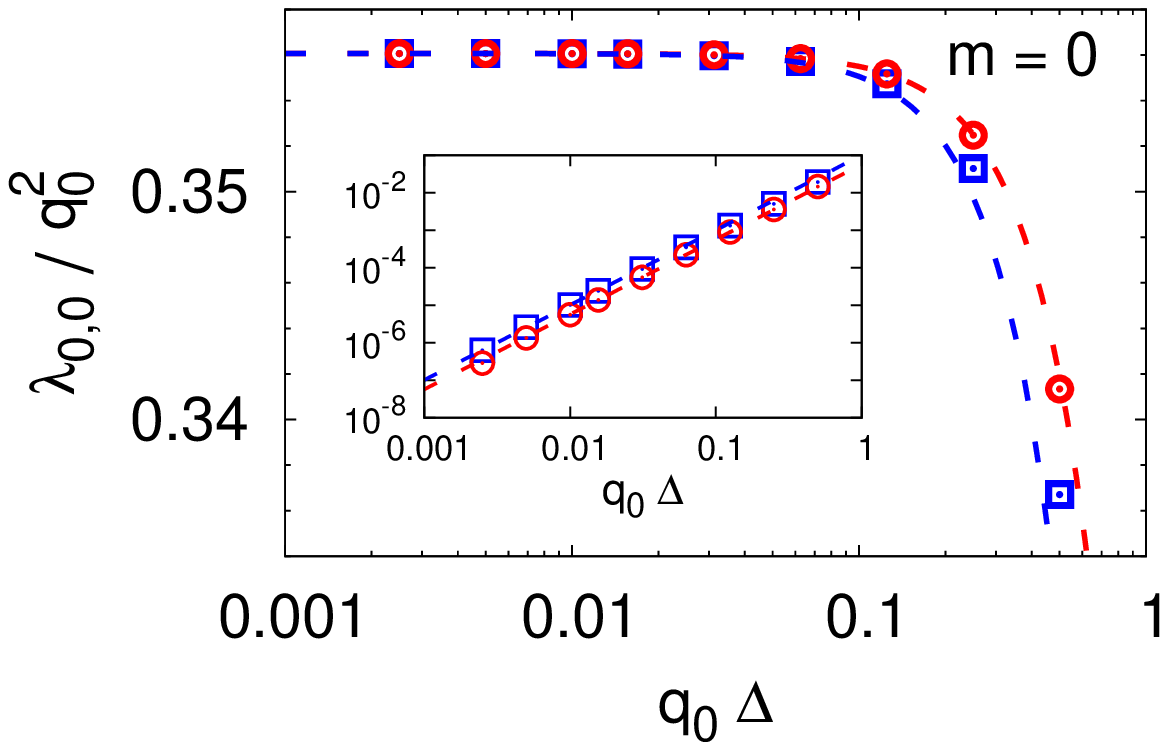}
\includegraphics[width=0.49\linewidth,angle=0]{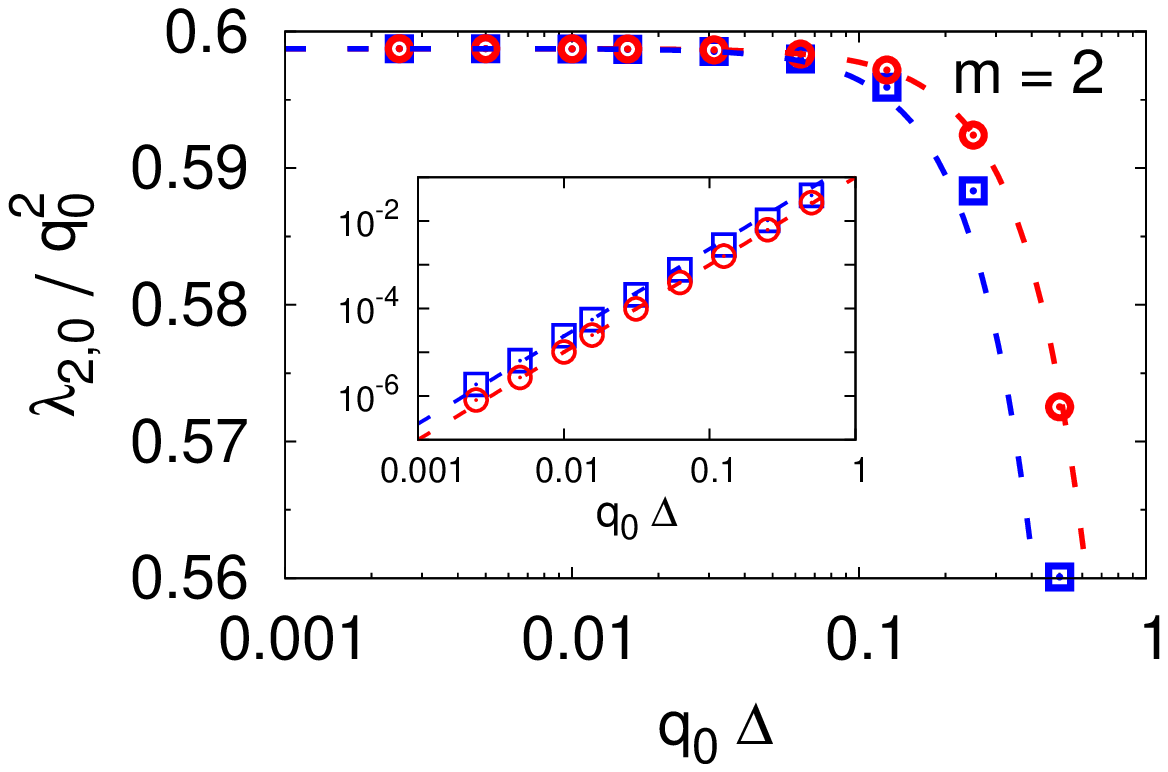}
\caption{Eigenvalue of the skyrmion fluctuation bound states $m=0$ (left) 
and $m=2$ (right) for $h=0.85$, obtained
with the improved (red circles) and simple (blue squares) discretizations,
\textit{versus} the discretization parameter $q_0\Delta$. 
The dashed lines represents the fits of~(\protect\ref{eq:fit2}) to the data in the
interval $q_0\Delta\in [0.0025,0.02]$. In all cases they are compatible with
second order convergence ($p=2$).
The insets display the effect of the discretization, defined as
the difference between the eigenvalue computed in the corresponding grid and the continuum limit,
estimated through the extrapolation $\lambda^{\mathrm{(c)}}$ (table~\protect\ref{tab:fit_sk}).
\label{fig:sk_bound}}
\end{figure}

The convergence rate with the improved discretization is also faster 
for the other two bound states ($m=0$ and $m=2$), as shown in Figs.~\ref{fig:sk_bound},
left and right, respectively.
The improvement, however, is not as dramatic as for the zero mode. The reason is that the
zero mode behaves in the vicinity of $r=0$ as $r^{1/2}$, while the other two bound states
behave as $r^{3/2}$. Indeeed, in both discretizations the convergence rate is described by
the exponent $p=2$ that corresponds to second order convergence. 
The results of fits of~(\ref{eq:fit2}) to the data are gathered in table~\ref{tab:fit_sk}.
Notice that the extrapolations to the continuum limit given by both discretizations agree to 
a high accuracy.
The insets in Figs.~\ref{fig:sk_bound} display the effect of the discretization, defined as
the difference between the eigenvalue computed in the corresponding grid and the continuum 
limit, estimated through the extrapolation $\lambda^{\mathrm{(c)}}$.

\section{Conclusions \label{sec:conc}}

The simplest central difference discretization of the second derivative of a function 
converges very slowly to the continuum limit at points where the function is singular 
in such a way that its fourth derivative becomes very large. This happens
in some instances of radial equations, depending on the coefficient of the $1/r^2$ term,
which usually represents the centrifugal potential. In such cases, the solution of the
radial equation through the diagonalization of the discretized equation converges to
the continuum limit at an extremely slow rate. This problem is caused by the singular
nature of the function and is not cured by a higher order discretization of the second 
derivative.
The convergence is extraordinarily accelerated if the $1/r^2$ term is discretized in 
such a way that the form of the continuum solution in the vicinity of the origin is 
exactly reproduced in the discretized problem. We call this finite differences 
scheme the \textit{centrifugal improved discretization}.

The centrifugal improved discretization is relevant in two dimensional problems, where
the radial function behaves as $r^\nu$ for $r\rightarrow 0$, with the exponent
$\nu$ non-integer. This is typical of two dimensional problems.
In three dimensional problems $\nu$ is always an integer if the
potential is regular, $\lim_{r\rightarrow 0} r^2V(r) = 0$, since in this case the $1/r^2$ 
term represents the centrifugal potential. However, for transition potentials, defined as 
those for which $\lim_{r\rightarrow 0} r^2V(r)=g$ is a real non-zero number \cite{Frank71}, 
the exponent $\nu$ is in general not an integer. In the limiting case $g\rightarrow -1/4$,
$\nu=1/2$. The method presented in this paper is thus very well suited for the numerical
study of transition potentials, which show very special theoretical features \cite{Frank71}.
The simplest transition potential, the inverse square potential, $V(r)=g/r^2$, has been 
extensively studied as it is exactly solvable (see for instance \cite{Camblong00}). 
Transition potentials are not just theoretically interesting, but have many physical 
applications. For instance, the potential due to a dipole is of inverse square type 
and has been applied long ago to the capture of electrons by polar molecules \cite{Levy67}.
Other experimental realizations of the inverse square potential and its associated 
phenomenology have also been reported \cite{Desfracois94,Denschlag98}.

For singular potentials, defined as those that are more singular than $1/r^2$,
the behaviour of the radial function at the origin is not determined by the $1/r^2$ term
but by the most singular term of the potential. For these cases, an improved finite differences
scheme analogous to the centrifugal improved discretization can be devised,
discretizing the potential in such a way that the form of the radial function as 
$r\rightarrow 0$ is exactly reproduced by the discretized solution.
This prescription will provide a powerful tool to study singular potentials, which are not just 
academic problems, since they appear in many physical systems. 
For instance, the potential interaction between one polar and one non-polar molecule is of $1/r^4$ type.

Finally, the centrifugal improved discretization can be also used to find numerically symmetric
solutions of nonlinear equations provided that the nonlinearity does not alter the singular behavior 
at the origin.
An instance is the Gross-Pitaevskii equation \cite{Gross61,Pitaevskii61} with cylindric symmetry, 
in which the nonlinear term in the radial equation has the form $u^3/r^{3/2}$, so that the most 
singular term at the origin is the centrifugal potential. 
A relaxation method can be used to solve the discretized nonlinear equation.

The authors acknowledge the Grant No. MAT2015-68200-C2-2-P from the Spanish Ministry of Economy and 
Competitiveness. This work was partially supported by the scientific JSPS Grant-in-Aid for Scientific 
Research (S) (Grant No. 25220803), and the MEXT program for promoting the enhancement of research
universities, and JSPS Core-to-Core Program, A. Advanced Research Networks.

\bibliographystyle{unsrt}
\bibliography{refs_radial}

\end{document}